\documentstyle[aps,preprint]{revtex}
\def\sbox#1{\mbox{\scriptsize #1}}
\def\T{\mbox{T}}
\def\R{\mbox{R}}
\def\L{\mbox{L}}

\def\sR{\sbox{R}}
\def\sL{\sbox{L}}
\def\H{{\cal H}}
\def\v#1{\mbox{\boldmath$#1$}}

\title
{Density Matrix Renormalization Group Method\\ for the Random Quantum
One-Dimensional Systems\\{\it - Application to the Random Spin-1/2
Antiferromagnetic Heisenberg Chain - }
}

\author
{Kazuo {\sc Hida}\footnote{e-mail: hida@th.phy.saitama-u.ac.jp}
}
\address
{
 Department of Physics, Faculty of Science,\\  Saitama University, Urawa,
Saitama 338
}

\date
{
 \today
}

\begin{document}

\maketitle

\begin{abstract}
The density matrix renormalization group method is generalized to one
dimensional random systems. Using this method, the energy gap distribution of
the spin-1/2 random antiferromagnetic Heisenberg chain is calculated. The
results are consistent with the predictions of the renormalization group theory
demonstrating the effectiveness of the present method in random systems. The
possible application of the present method to other random systems is
discussed.
\end{abstract}

\newpage
\section{Introduction}
Low dimensional quantum spin systems have been studied intensively in this
decade particularly stimulated by its relevance to high $T_c$
superconductivity. Compared to the regular systems, however, the random quantum
spin systems are less studied, while their classical counterparts are also
intensively studied in the context of spin glass problem.

One of the pioneering works on the quantum random antiferromagnet was done by
Blatt and Lee\cite{bl1} who proposed the random singlet phase. This model was
further studied by means of the renormailztion group method by DasGupta and
Ma\cite{dgm1}, Hirsch and Jos\'e\cite{hj1} and Fisher\cite{df1} for the
one-dimensional case. This method was recently applied to the random chain with
ferromagnetic and antiferromagnetic interaction by Westerberg {\it et
al}.\cite{wfsl1,furu,furul} and to the one-dimensional random transverse Ising
model by Fisher.\cite{df2}

On the other hand, Young and Rieger\cite{yr1} recently analyzed the
one-dimensional random transverse Ising model numerically up to the chain
length $N=128$ taking advantage of the fact that this model is mapped onto the
random {\it free} spinless fermion model. Their results are in good agreement
with the renormalization group theory.\cite{df2} However, the numerical study
of the ground state of the strongly interacting random systems still remains
rather limited. The conventional numerical approach such as exact
diagonalization by Lanzcos algorithm requires considerable computational time
and memory even for a single system if the system size is large. Such
calculation becomes even harder for the random systems for which the average
over {\it many large} samples must be taken.

The density matrix renormalization group (DMRG) method initiated by
White\cite{srw1,srw2} is a promising candidate to overcome this difficulty. The
major advantage of this method is that it enables one to calculate the low
lying states of very large one-dimensional systems quite accurately with
relatively small computational time and memory. However, this method is
originally developed for the regular uniform systems taking advantage of the
uniformness of the system. It is the purpose of the present work to introduce
the algorithm to apply this method to random chains. As an example of the
application of the present method, we also calculate the energy gap
distribution of the random spin-1/2 antiferromagnetic Heisenberg chain which
shows good agreement with the renormalization group theory.\cite{dgm1,hj1,df1}

\section{DMRG Method for the Random Chain}

The usual procedure of the infinite size algorithm of the DMRG calculation for
the uniform system is as follows,

\begin{enumerate}
\item Prepare the left and right blocks $\L_{N/2}$ and $\R_{N/2}$ where $N/2$
is the number of sites in each block. To start with, choose $N=4$. In what
follows, $N$ is always assumed to be even.

\item Construct the Hamiltonian of the superblock $\T_{N}$ connecting the right
end of $\L_{N/2}$ and and left end of $\R_{N/2}$. Diagonalize it to obtain its
low lying states (called target states).
\item  Construct the density matrices $\rho_{\sL}$ ($\rho_{\sR})$ for
$\L_{N/2}$ ($\R_{N/2}$) by tracing out the states of $\R_{N/2}$ ($\L_{N/2}$)
from the target states.

\item Diagonalize the density matrices $\rho_{\sL}$ and $\rho_{\sR}$ and choose
$m$ important eigenstates of $\L_{N/2}$ and $\R_{N/2}$.

\item Add a unit cell at the right end of $\L_{N/2}$ and the left end of
$\R_{N/2}$ to construct $\L_{N/2+1}$ and $\R_{N/2+1}$ keeping only $m$ states
chosen in the preceding step.

\item Setting $N+2 \rightarrow N$, go to step 2 and repeat.
\end{enumerate}
This procedure is schematically shown in Fig. \ref{fig1}.

White\cite{srw1,srw2} has shown that this method allows to calculate the low
energy states of very large systems quite accurately, if we pick up appropiate
number of states in both left and right blocks.

The essential point which requires the uniformity of the system is the
assumption that the important states in $\L_{N/2}$ and $\R_{N/2}$ are also
important in $\L_{N/2+1}$ and $\R_{N/2+1}$. This is garanteed because the
enviroment of the boundary spins hardly changes by the addition of the extra
spins between  $\L_{N/2}$ and $\R_{N/2}$  as far as the system is uniform.

This means that we have to pick up the states in  $\L_{N/2}$ and $\R_{N/2}$
which satisfies the boundary condition that {\it it is connected with the
neighbouring block} in general.  However, in the random system, all bonds are
inequivalent. Thefore if we add spins between the two blocks, the boundary
condition for the block changes and the important states in $\L_{N/2}$ and
$\R_{N/2}$ are no more important in  $\L_{N/2+1}$ and $\R_{N/2+1}$. This is the
difficulty in extending the DMRG method to the random systems.

To circumvent this difficulty, we prepare all the blocks from the beginning  so
that they provide the proper enviroment for each other in each step and let
them grow in parallel keeping the boundary conditions for each block almost
unchanged during the course of the calculation.

The calculation proceeds as follows:

\begin{enumerate}
\item Prepare a large enough even membered sample $[1:N]$ which consists of the
sites $i=1,...N$. Here, $[i:j]$ denote the superblock consisting of the $i,i+1,
..,j$-th sites.
\item Cut out all four-site superblocks $[2i-1:2i+2]$ ($i=1,N/2-1)$ and
diagonalize their Hamiltonian to get their target states.
\item By the procdures 3 to 4 for the regular system, find the important states
of the blocks $[2i-1:2i)$ and $(2i+1:2i+2]$ for all $i$. Here the ends
connected with the neighbouring blocks are denoted by ( and ), while [ and ]
denote the open ends.
\item Add $2i+1$-th spin to the right end of $[2i-1:2i)$ and $2i+2$-th spin to
the left end of $(2i+3:2i+4]$. Diagonalize the Hamiltonian of the superblock
$[2i-1:2i+4]$ constructed from the important states of  $[2i-1:2i)$ and
$(2i+3:2i+4]$ in addition to all the states of the $2i+1$-th and $2i+2$-th
spins.  By the procedure 3 to 4 for the regular system, find the important
states of the blocks $[2i-1:2i+1)$ and $(2i+2:2i+4]$ for all $i$.

\item Repeat until the superblock $[1,N]$ is diagonalized.
\end{enumerate}
This is schematically shown in Fig. \ref{fig2} for $N=10$.
It should be emphasized the above procedure must be performed in parallel for
all $i$ on each step. Therefore we obtain the low lying states of even membered
subsystems which are superblocks in each step as by-product. If we pick up only
independent subsystems, the data for {\it many small} systems are obtained
during the course of the calculation of a large single system.

This algorithm corresponds to the 'infinite size method' in the regular case.
Therefore the eigenstates obtained in each step are only approximate. To
improve the accuracy, we can also include the finite size iteration whose
algorithm is an obvious extension of the regular case.

\section{Application to the Random Spin-1/2 Antiferromagnetic Chain}

As an example, we have applied this method to the random spin-1/2 Heisenberg
chain whose Hamiltonian is given by,
\begin{equation}
\label{eq:ham}
\H =\sum_{i=1}^{N} 2J_i\v{S}_{i}\v{S}_{i+1},\ \ \mid \v{S}_{i}\mid = 1/2,
\end{equation}
where $J_i$ takes the random positive values. Here we assume the distribution

\begin{equation}
J_i = \left\{\begin{array}{ll}
J & \mbox{with probability}\ \ p, \\
J' & \mbox{with probability}\ \ 1-p.
\end{array}\right.
\end{equation}

The number of prepared samples with $N=48$ ranged from 866 to 920. We have
studied the cases $p=0.3, 0.5$ and 0.7 with $J=1$ and $J'=0.5$. For all these
samples, the DMRG procedure is carried out. During the growth of the system, we
have obtained the energy gap for smaller systems with $N < 48$. The number $m$
of the states kept in each step is 60. We have also checked the accuarcy taking
$m=80$ and 100 for several samples. For all checked samples, we find $m=60$ is
sufficient without finite size iterations.

According to the renormalization group theory,\cite{dgm1,hj1,df1} the logarithm
of the characteristic energy $\Omega$ of the $N$-membered system scales with
$\sqrt{N}$. Roughly speaking, this energy scale can be interpreted as a random
{\it product} of the energy scales of the subsystems. Therefore $\ln\Omega$ is
a good statistical variable rather than $\Omega$ itself. In addition, the
renormalization group theory predicts that not only the root mean square
deviation of $\ln\Omega$ but also their overall distribution function is scaled
by $\sqrt{N}$.

As a characteristic energy of each cluster, we have calculated the lowest
energy gap $\Delta$. Figure \ref{fig3} shows the system size dependence of the
average $<\ln \Delta>$ and root mean square deviation $\sigma \equiv
\sqrt{<(\ln \Delta-<\ln \Delta>)^2>}$.

They are well fitted by the linear function of $\sqrt{N}$. In addition, the
ratio $<\ln \Delta>/\sigma$ remained almost equal to 2 irrespective of $N$ and
$p$ as shown in \ref{fig4}. This suggests that the distribution is scaled by a
single variable $\Gamma \equiv \ln (\Delta/\Delta_0)/\sqrt{N}$ as expected from
the renormalization group theory where  $\ln \Delta_0$ is determined by the
linear extrapolation of $<\ln \Delta>$ to $\sqrt{N} \rightarrow 0$.

Based on this observation, the logarithm of the distribution function
$P(\Gamma)$ is plotted against $\Gamma^2$ in Fig. \ref{fig5}. It follows the
universal curve irrespective of the system size in accordance with the
renormalization group prediction. For small $\Delta$ ( large $\Gamma$ ),  $\ln
P(\Gamma)$ is linear to  $\Gamma^2$ . This is also in agreement with the
prediction of the renormalization group theory.\cite{df1} It should be remarked
that the renormalization group calculation becomes more and more accurate for
small energy scale. Thus this agreement confirms that our algorithm of the DMRG
for the random system works sufficiently well.

All the data from Fig. \ref{fig3} to \ref{fig5} depend weakly on $p$. It is not
clear whether this actually implies the breakdown of universality with respect
to $p$. From Fig. \ref{fig5}, the major difference comes from the samples with
very small $\Delta$ for which the numerical accuarcy is relatively poor. In
addition, the number of samples having such small gaps is small and statistical
error is also expected to be large in this regime.

\section{Summary and Discussion}

The algorithm to apply the DMRG method to the random one-dimensional systems is
presented. It is demonstrated that the results of the renormalization group
theory for the random spin-1/2 antiferromagnetic Heisenberg chains are
reproduced using our method.

As discussed in the introduction, the random quantum spin chain is attracting
the renewed interest quite recently. It is predicted that the random Heisenberg
chain with ferromagnetic and antiferromagnetic interactions belongs to the
universality class different from random spin-1/2 antiferromagnetic Heisenberg
chain.\cite{wfsl1}. The DMRG study of this model is now being carried out.

The present algorithm is also useful to the study of the systems with large
unit cell. In the conventional DMRG, two unit cells must be added on each step
and therefore the effective Hamiltonian matrix of the superblock becomes huge
and hardly tractable. On the contrary, in the present method, we need to add
only 2 spins on each steps even if the unit cell is large. Therefore we can
keep the size of the Hamiltonian matrix modest. For example, we may apply this
method to the trimerized system\cite{khtr1,ajiro1,oktr1}.

Application of the present method is not limited to the quantum spin system.
The itinerant electron systems should provide wide possibility of application.
For example, the interplay between the Anderson localization and electron
correlation or the quantum effect on the randomly pinned charge density wave
remain to be investigated even in the one-dimensional systems.\cite{saso1}

Recently, Nishino and Okunishi\cite{no1} have introduced the product wave
function renormalization group method based on the observation of \"Ostlund and
Rommer\cite{or1} that the DMRG wave function is expressed as a product of
matrices. This method is much faster than the conventional DMRG method, because
it can skip the Lanzcos diagonalization of superblocks. If this method can be
incorporated into the present algorithm, the computational efficiency will be
much improved. This is left for future study.

The numerical calculation has been performed using the FACOM VPP500 at the
Supercomputer Center, Institute for Solid State Physics, University of Tokyo,
the HITAC S3800 at the Computer Center, University of Tokyo and the HITAC
S820/15 at the Information Processing Center, Saitama University.  This work is
supported by the Grant-in-Aid for Scientific Research from the Ministry of
Education, Science and Culture.

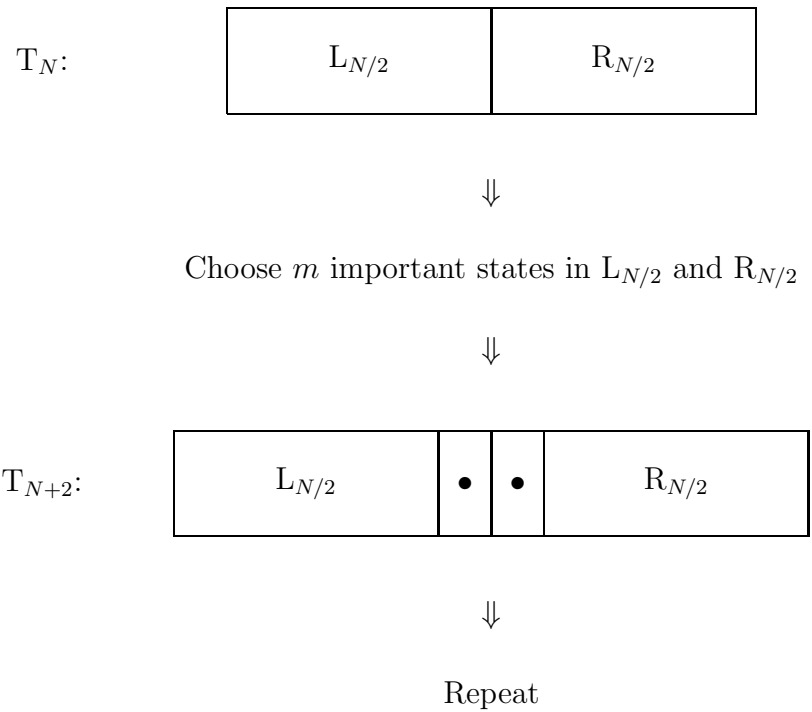
\begin{figure}
\caption{The schematic procedure of the DMRG calculation for the uniform
systems.}
\label{fig1}
\end{figure}
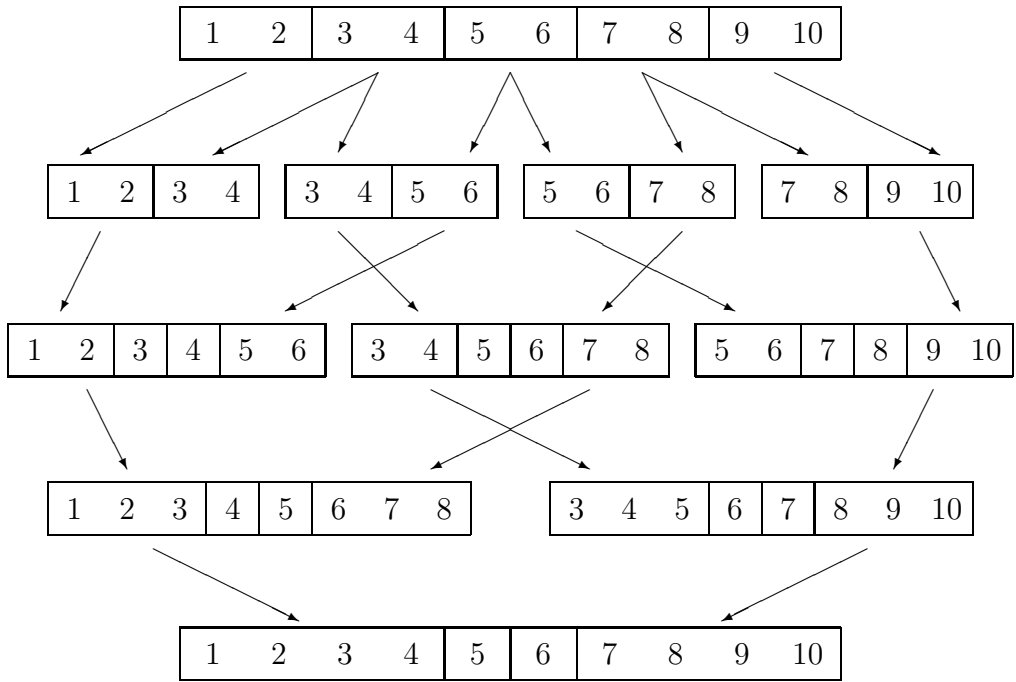
\begin{figure}
\caption{The schematic procedure of the DMRG calculation for the random systems
for $N=10$.}
\label{fig2}
\end{figure}
\begin{figure}
\caption{The system size dependence of the average $<\ln \Delta>$ and  root
mean square deviation $\sigma$ for $p=0.3$ ($\circ$ ; solid line), 0.5
($\bullet$; dotted line)  and 0.7 ($\Box$; broken line) with $J=1$ and
$J'=0.5$.}
\label{fig3}
\end{figure}
\begin{figure}
\caption{The relation between the average $<\ln \Delta>$ and root mean square
deviation $\sigma$ for  for $p=0.3$ ($\circ$ ; solid line), 0.5 ($\bullet$ ;
dotted line)  and 0.7 ($\Box$ ; broken line) with $J=1$ and $J'=0.5$.}
\label{fig4}
\end{figure}
\begin{figure}
\caption{The energy gap distribution  $P(\Gamma)$ plotted against $\Gamma^2
(\equiv \ln^2(\Delta E/\Delta_0)/N)$ for (a) $p=0.3$, (b) $p=0.5$ and (c)
$p=0.7$ with $J=1$ and $J'=0.5$. }
\label{fig5}
\end{figure}

\newpage

\begin{picture}(600,350)(0,50)
\put(100,330){\line(1,0){200}}
\put(100,330){\line(0,1){40}}
\put(100,370){\line(1,0){200}}
\put(300,330){\line(0,1){40}}
\put(200,330){\line(0,1){40}}
\put(150,350){\makebox(0,0)[c]{$\mbox{L}_{N/2}$}}
\put(250,350){\makebox(0,0)[c]{$\mbox{R}_{N/2}$}}
\put(30,350){\makebox(0,0)[c]{$\mbox{T}_{N}$:}}
\put(200,300){\makebox(0,0)[c]{$\Downarrow$}}
\put(200,270){\makebox(0,0)[c]{Choose $m$ important states in $\mbox{L}_{N/2}$
 and $\mbox{R}_{N/2}$}}
\put(200,240){\makebox(0,0)[c]{$\Downarrow$}}
\put(80,170){\line(1,0){240}}
\put(80,170){\line(0,1){40}}
\put(80,210){\line(1,0){240}}
\put(220,170){\line(0,1){40}}
\put(200,170){\line(0,1){40}}
\put(180,170){\line(0,1){40}}
\put(320,170){\line(0,1){40}}
\put(130,190){\makebox(0,0)[c]{$\mbox{L}_{N/2}$}}
\put(270,190){\makebox(0,0)[c]{$\mbox{R}_{N/2}$}}
\put(30,190){\makebox(0,0)[c]{$\mbox{T}_{N+2}$:}}
\put(190,190){\makebox(0,0)[c]{$\bullet$}}
\put(210,190){\makebox(0,0)[c]{$\bullet$}}
\put(200,140){\makebox(0,0)[c]{$\Downarrow$}}
\put(200,110){\makebox(0,0)[c]{Repeat}}
\put(200,0){\makebox(0,0)[c]{FIG. 1}}
\end{picture}

\newpage
\unitlength 0.5pt
\begin{picture}(600,550)(50,-120)
\put(200,350){\line(1,0){500}}
\put(200,390){\line(1,0){500}}
\put(200,350){\line(0,1){40}}
\put(300,350){\line(0,1){40}}
\put(400,350){\line(0,1){40}}
\put(500,350){\line(0,1){40}}
\put(600,350){\line(0,1){40}}
\put(700,350){\line(0,1){40}}
\put(225,370){\makebox(0,0)[c]{1}}
\put(275,370){\makebox(0,0)[c]{2}}
\put(325,370){\makebox(0,0)[c]{3}}
\put(375,370){\makebox(0,0)[c]{4}}
\put(425,370){\makebox(0,0)[c]{5}}
\put(475,370){\makebox(0,0)[c]{6}}
\put(525,370){\makebox(0,0)[c]{7}}
\put(575,370){\makebox(0,0)[c]{8}}
\put(625,370){\makebox(0,0)[c]{9}}
\put(675,370){\makebox(0,0)[c]{10}}

\put(250,340){\vector(-2,-1){125}}
\put(350,340){\vector(-2,-1){125}}
\put(350,340){\vector(-1,-2){30}}
\put(450,340){\vector(-1,-2){30}}
\put(450,340){\vector(1,-2){30}}
\put(650,340){\vector(2,-1){125}}
\put(550,340){\vector(2,-1){125}}
\put(550,340){\vector(1,-2){30}}

\put(100,270){\line(1,0){160}}
\put(100,230){\line(1,0){160}}
\put(280,270){\line(1,0){160}}
\put(280,230){\line(1,0){160}}
\put(460,270){\line(1,0){160}}
\put(460,230){\line(1,0){160}}
\put(640,270){\line(1,0){160}}
\put(640,230){\line(1,0){160}}

\put(100,230){\line(0,1){40}}
\put(180,230){\line(0,1){40}}
\put(260,230){\line(0,1){40}}

\put(280,230){\line(0,1){40}}
\put(360,230){\line(0,1){40}}
\put(440,230){\line(0,1){40}}

\put(460,230){\line(0,1){40}}
\put(540,230){\line(0,1){40}}
\put(620,230){\line(0,1){40}}

\put(640,230){\line(0,1){40}}
\put(720,230){\line(0,1){40}}
\put(800,230){\line(0,1){40}}

\put(120,250){\makebox(0,0)[c]{1}}
\put(160,250){\makebox(0,0)[c]{2}}
\put(200,250){\makebox(0,0)[c]{3}}
\put(240,250){\makebox(0,0)[c]{4}}

\put(300,250){\makebox(0,0)[c]{3}}
\put(340,250){\makebox(0,0)[c]{4}}
\put(380,250){\makebox(0,0)[c]{5}}
\put(420,250){\makebox(0,0)[c]{6}}

\put(480,250){\makebox(0,0)[c]{5}}
\put(520,250){\makebox(0,0)[c]{6}}
\put(560,250){\makebox(0,0)[c]{7}}
\put(600,250){\makebox(0,0)[c]{8}}

\put(660,250){\makebox(0,0)[c]{7}}
\put(700,250){\makebox(0,0)[c]{8}}
\put(740,250){\makebox(0,0)[c]{9}}
\put(780,250){\makebox(0,0)[c]{10}}

\put(140,220){\vector(-1,-2){30}}
\put(320,220){\vector(1,-1){60}}
\put(400,220){\vector(-2,-1){120}}
\put(500,220){\vector(2,-1){120}}
\put(580,220){\vector(-1,-1){60}}
\put(760,220){\vector(1,-2){30}}

\put(70,150){\line(1,0){240}}
\put(70,110){\line(1,0){240}}
\put(330,150){\line(1,0){240}}
\put(330,110){\line(1,0){240}}
\put(590,150){\line(1,0){240}}
\put(590,110){\line(1,0){240}}

\put(70,110){\line(0,1){40}}
\put(150,110){\line(0,1){40}}
\put(190,110){\line(0,1){40}}
\put(230,110){\line(0,1){40}}
\put(310,110){\line(0,1){40}}

\put(330,110){\line(0,1){40}}
\put(410,110){\line(0,1){40}}
\put(450,110){\line(0,1){40}}
\put(490,110){\line(0,1){40}}
\put(570,110){\line(0,1){40}}

\put(590,110){\line(0,1){40}}
\put(670,110){\line(0,1){40}}
\put(710,110){\line(0,1){40}}
\put(750,110){\line(0,1){40}}
\put(830,110){\line(0,1){40}}

\put(90,130){\makebox(0,0)[c]{1}}
\put(130,130){\makebox(0,0)[c]{2}}
\put(170,130){\makebox(0,0)[c]{3}}
\put(210,130){\makebox(0,0)[c]{4}}
\put(250,130){\makebox(0,0)[c]{5}}
\put(290,130){\makebox(0,0)[c]{6}}

\put(350,130){\makebox(0,0)[c]{3}}
\put(390,130){\makebox(0,0)[c]{4}}
\put(430,130){\makebox(0,0)[c]{5}}
\put(470,130){\makebox(0,0)[c]{6}}
\put(510,130){\makebox(0,0)[c]{7}}
\put(550,130){\makebox(0,0)[c]{8}}

\put(610,130){\makebox(0,0)[c]{5}}
\put(650,130){\makebox(0,0)[c]{6}}
\put(690,130){\makebox(0,0)[c]{7}}
\put(730,130){\makebox(0,0)[c]{8}}
\put(770,130){\makebox(0,0)[c]{9}}
\put(810,130){\makebox(0,0)[c]{10}}

\put(130,100){\vector(1,-2){30}}
\put(390,100){\vector(2,-1){120}}
\put(510,100){\vector(-2,-1){120}}
\put(770,100){\vector(-1,-2){30}}

\put(100,30){\line(1,0){320}}
\put(100,-10){\line(1,0){320}}
\put(480,30){\line(1,0){320}}
\put(480,-10){\line(1,0){320}}

\put(100,-10){\line(0,1){40}}
\put(220,-10){\line(0,1){40}}
\put(260,-10){\line(0,1){40}}
\put(300,-10){\line(0,1){40}}
\put(420,-10){\line(0,1){40}}

\put(480,-10){\line(0,1){40}}
\put(600,-10){\line(0,1){40}}
\put(640,-10){\line(0,1){40}}
\put(680,-10){\line(0,1){40}}
\put(800,-10){\line(0,1){40}}

\put(120,10){\makebox(0,0)[c]{1}}
\put(160,10){\makebox(0,0)[c]{2}}
\put(200,10){\makebox(0,0)[c]{3}}
\put(240,10){\makebox(0,0)[c]{4}}
\put(280,10){\makebox(0,0)[c]{5}}
\put(320,10){\makebox(0,0)[c]{6}}
\put(360,10){\makebox(0,0)[c]{7}}
\put(400,10){\makebox(0,0)[c]{8}}

\put(500,10){\makebox(0,0)[c]{3}}
\put(540,10){\makebox(0,0)[c]{4}}
\put(580,10){\makebox(0,0)[c]{5}}
\put(620,10){\makebox(0,0)[c]{6}}
\put(660,10){\makebox(0,0)[c]{7}}
\put(700,10){\makebox(0,0)[c]{8}}
\put(740,10){\makebox(0,0)[c]{9}}
\put(780,10){\makebox(0,0)[c]{10}}

\put(180,-20){\vector(2,-1){110}}
\put(720,-20){\vector(-2,-1){110}}

\put(200,-120){\line(1,0){500}}
\put(200,-80){\line(1,0){500}}
\put(200,-120){\line(0,1){40}}
\put(400,-120){\line(0,1){40}}
\put(450,-120){\line(0,1){40}}
\put(500,-120){\line(0,1){40}}
\put(700,-120){\line(0,1){40}}
\put(225,-100){\makebox(0,0)[c]{1}}
\put(275,-100){\makebox(0,0)[c]{2}}
\put(325,-100){\makebox(0,0)[c]{3}}
\put(375,-100){\makebox(0,0)[c]{4}}
\put(425,-100){\makebox(0,0)[c]{5}}
\put(475,-100){\makebox(0,0)[c]{6}}
\put(525,-100){\makebox(0,0)[c]{7}}
\put(575,-100){\makebox(0,0)[c]{8}}
\put(625,-100){\makebox(0,0)[c]{9}}
\put(675,-100){\makebox(0,0)[c]{10}}
\put(450,-200){\makebox(0,0)[c]{FIG. 2}}

\end{picture}


\begin{references}
\bibitem{bl1} R. N. Bhatt and P. A. Lee, Phys. Rev. Lett. {\bf 48}
 (1982) 344.
\bibitem{dgm1} C. Dasgupta and S.-k. Ma, Phys. Rev. B {\bf 22} (1980) 1305;
 S.-k. Ma, C. Dasgupta, and C. K. Hu, Phys. Rev. Lett. {\bf 43} (1979) 1434.
\bibitem{hj1}  J. E. Hirsch and J. V. Jos\'e, Phys. Rev. B {\bf 22} (1980) 5339
; J. E. Hirsch, {\it ibid.} {\bf 22} (1980) 5355.
\bibitem{df1} D. S. Fisher, Phys. Rev. B {\bf 50} (1994) 3799
{}.
\bibitem{wfsl1} E. Westerberg, A. Furusaki, M. Sigrist, and P. A. Lee, Phys.
Rev. Lett. {\bf 75} (1995) 4302.
\bibitem{furu} A. Furusaki, M. Sigrist, P. A. Lee, K. Tanaka, and
 N. Nagaosa, Phys. Rev. Lett. {\bf 73} (1994) 2622.
\bibitem{furul}  A. Furusaki, M. Sigrist, E. Westerberg, P. A. Lee, K. B.
Tanaka, and N. Nagaosa, preprint, cond-mat.9507083 (1995).
\bibitem{df2} D. S. Fisher, Phys. Rev. Lett. {\bf 69} (1992) 534; Phys. Rev. B
{\bf51} (1995) 6411.
\bibitem{yr1} A. P. Young and H. Rieger: preprint cond-mat.9510027 (1995)
\bibitem{srw1}   S. R. White: Phys. Rev. Lett. {\bf 69} 2863 (1992).
\bibitem{srw2}   S. R. White: Phys. Rev. {\bf B48} 10345 (1993).
\bibitem{khtr1} K. Hida: J. Phys. Soc. Jpn. {\bf 63} (1994) 2359.
\bibitem{ajiro1} Y. Ajiro, T. Asano, T. Inami, H. Aruga-Katori and T. Goto: J.
Phys. Soc. Jpn. {\bf 63} (1994) 859.
\bibitem{oktr1} K. Okamoto: preprint cond-mat.9510070 (1995).
\bibitem{saso1} For example T. Saso and Y. Suzumura: J. Phys. Soc. Jpn. (1986)
3151.
\bibitem{no1} T. Nishino and K. Okunishi: J. Phys. Soc. Jpn. (1996) to appear.
\bibitem{or1} S. \"Ostlund and S. Rommer: preprint cond-mat.9503107
\end{references}
\end{document}